\documentclass[times, final]{elsarticle}
\usepackage[utf8]{inputenc}
\usepackage{amsmath}
\usepackage{amsfonts}
\usepackage{amssymb}
\usepackage{mathrsfs}
\usepackage{subfigure}
\usepackage[hidelinks]{hyperref}
\usepackage{outlines}
\usepackage{tikz}
\usepackage{pgfplots}
\usepackage{wrapfig}

\biboptions{numbers, sort&compress}

\newcommand{\bv}{\mathbf{v}}

\newcommand{\bE}{\mathbf{E}}
\newcommand{\bB}{\mathbf{B}}

\newcommand{\tn}{\tilde{n}}
\newcommand{\tT}{\tilde{T}}
\newcommand{\tB}{\tilde{B}}
\newcommand{\tv}{\tilde{v}}

\usepackage{tensor}

\newcommand{\lp}{\left(}
\newcommand{\rp}{\right)}

\DeclareFontFamily{OMS}{oasy}{\skewchar\font48 }
\DeclareFontShape{OMS}{oasy}{m}{n}{%
	<-5.5> oasy5     <5.5-6.5> oasy6
	<6.5-7.5> oasy7     <7.5-8.5> oasy8
	<8.5-9.5> oasy9     <9.5->  oasy10
}{}
\DeclareFontShape{OMS}{oasy}{b}{n}{%
	<-6> oabsy5
	<6-8> oabsy7
	<8->  oabsy10
}{}
\DeclareSymbolFont{oasy}{OMS}{oasy}{m}{n}
\SetSymbolFont{oasy}{bold}{OMS}{oasy}{b}{n}

\DeclareMathSymbol{\smallleftarrow}     {\mathrel}{oasy}{"20}
\DeclareMathSymbol{\smallrightarrow}    {\mathrel}{oasy}{"21}
\DeclareMathSymbol{\smallleftrightarrow}{\mathrel}{oasy}{"24}

\begin{document}

\title{MITNS: Multiple-Ion Transport Numerical Solver for Magnetized Plasmas}

\fntext[fn1]{Co-first authors}

\date{\today}
\author{E.J. Kolmes\fnref{fn1}}
\ead{ekolmes@princeton.edu}
\author{I.E. Ochs\fnref{fn1}}
\ead{iochs@princeton.edu}
\author{N.J. Fisch}
\address{Department of Astrophysical Sciences, Princeton University, Princeton, NJ 08544, USA}

\begin{abstract}
MITNS (Multiple-Ion Transport Numerical Solver) is a new numerical tool designed to perform 1D simulations of classical cross-field transport in magnetized plasmas. Its detailed treatment of multi-species effects makes it a unique tool in the field. We describe the physical model it simulates, as well as its numerical implementation and performance. 
\end{abstract}

\maketitle

\begin{small}
\noindent
{\em Program Title:} MITNS (Multiple-Ion Transport Numerical Solver) \\
{\em Licensing provisions:} MIT \\
{\em Programming language:} C++, with Python wrapper \\
{\em Nature of problem:} 
Classical transport of multiple-species plasma across a magnetic field. This includes the collisional transport of particles, momentum, and heat. These quantities are tracked separately for each particle species. Both ion-ion and ion-electron interactions are included, as is the evolution of the magnetic field. 
\\
{\em Solution method:} The system of PDEs is decomposed into a large system of coupled ODEs. The code uses finite-volume discretization for space. Time integration is done using any of three timestepping methods, including Adams-Moulton and Backwards Differentiation Formula schemes from the CVODE package [1, 2]. 

\end{small}

\section{Introduction}

MITNS is a 1D multiple-fluid simulation code designed to study classical cross-field transport physics, with a particular focus on plasmas containing multiple ion species. 
Multiple-species cross-field transport problems are important across a wide range of plasma applications, including nuclear fusion devices like tokamaks \cite{Hirshman1981, Redi1991, Wade2000, Dux2004}, stellarators \cite{Braun2010, Helander2017, Newton2017}, various pinch configurations \cite{Shumlak1995, Shumlak2003, Shumlak2012, Slutz2012, HarveyThompson2018}, and non-fusion technologies like plasma mass filters \cite{Bonnevier1966, Lehnert1971, Krishnan1983, Zweben2018, Gueroult2018}. MITNS is designed specifically to simulate classical transport, which means that it is not designed to study regimes controlled by ``anomalous" transport (e.g., due to turbulence) or the neoclassical effects that can arise in toroidal systems. 
Hence, MITNS is not primarily intended for tokamak or stellarator applications. 

There are other plasma simulation codes that include related physics in one form or another. 
For instance, the GBS code simulates the Braginskii two-fluid equations (that is, for a single ion fluid and electrons) \cite{Ricci2012}. Other authors have worked with multiple-fluid simulations that include neutrals and one ion species \cite{Leake2014, AlvarezLaguna2016}. 
B2.5, UEDGE, and EDGE2D/U all use $N$-fluid models to track different ion species' densities and momenta independently; these codes assume that all ion species share a single temperature profile and their physical models for cross-field transport are anomalous rather than classical \cite{Rognlien1994, Simonini1994, Radford1996, Braams1996, Wiesen2015, Rognlien2018}. 
There has also been significant computational work using $N$-fluid models to simulate unmagnetized multiple-ion plasmas \cite{Rambo1994, Rambo1995, Ghosh2019}. 
To the authors' knowledge, there is no established code designed to simulate classical $N$-fluid cross-field transport, including independent densities, velocities, and temperatures, for an arbitrary number of species.

MITNS does not model complex magnetic geometries, like a tokamak or a stellarator. 
Rather, it is designed to capture, isolate, and aid in the understanding of the fundamental physics of classical cross-field transport in mixed-species plasmas. 
Thus it models a deliberately simple 1D geometry, without effects like transport parallel to the magnetic field or interactions with plasma-facing components.

This paper describes in detail the MITNS code: its model, its assumptions, and its numerical properties. As other studies begin to be released that rely on MITNS \cite{MlodikHeatPumparXiv}, this paper serves as a more detailed description of the code than would gracefully fit elsewhere. That includes a description of the physical model as well as its underlying caveats and domains of applicability. Sections~\ref{sec:model} and \ref{sec:dimensionless} describe the physical model that MITNS simulates. Section \ref{sec:scaling} discusses the ways in which the code allows different physical phenomena to be turned off or scaled up and down. These details may be intrinsically interesting to others working on similar numerical problems; in particular, the implementation of the scalable thermal conductivity involves nontrivial physics and could be applicable to other codes. 

MITNS uses finite-volume spatial discretization. It can perform time integration using any of three schemes: fourth-order Runge-Kutta (RK4), Adams-Moulton (AM), and Backwards Differentiation Formula (BDF). It relies on components of the SUNDIALS suite, including some data structures and implementations of the AM and BDF time integration \cite{Hindmarsh2005, Cohen1996}. The implementation of the code is described in Section~\ref{sec:implementation} and its performance is discussed in Section~\ref{sec:performance}. 

\section{Physical Model Equations} \label{sec:model}

\begin{figure}
	\centering
		\begin{tikzpicture} [
		declare function={a(\x)=\x;},
		declare function={b(\x)=0.5*\x-1;}
		]
		
		\draw[very thick, black] (0,0) rectangle (4,4);
		
		\draw[->, very thick, blue] (4.5,3.) -- (4.5,1.);
		\node at (4.8, 2.) {\Large \color{blue} $g$}; 
		
		\draw[<->, very thick, red] (-0.5, 1.) -- (-0.5, 3.); 
		\node at (-1.7, 2.) {\Large \color{red}{gradients}}; 
		
		\draw[<->, very thick, violet] (.5,-.5) -- (3.5,-.5);
		\node at (2., -1.) {\Large \color{violet}{drift flow}}; 
		
		\draw[->, very thick, black] (.5,.5) -- (1.5,.5);
		\draw[->, very thick, black] (.5,.5) -- (.5,1.5);
		\draw[->, very thick, black] (.5,.5) -- (1.1,0.9);
		
		\node at (1.8,.5) {\Large $\hat y$};
		\node at (.5,1.8) {\Large $\hat x$};
		\node at (1.3,1.1) {\Large $\hat z$}; 
		
		\draw[very thick, black] (3,3.5) circle (.25);
		\draw[very thick, black] (3+.25*.707,3.5+.25*.707) -- (3-.25*.707,3.5-.25*.707);
		\draw[very thick, black] (3-.25*.707,3.5+.25*.707) -- (3+.25*.707,3.5-.25*.707);
		
		\node at (3.6, 3.5) {\Large $\mathbf{B}$};
		
		\end{tikzpicture}
	\caption{This schematic shows the basic geometry and coordinates used in MITNS. } \label{fig:geometrySchematic}
\end{figure}
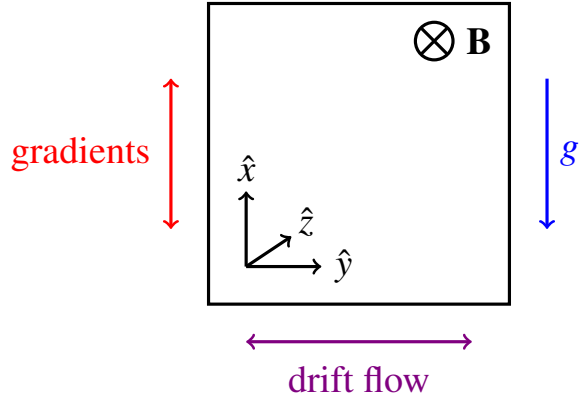

MITNS simulates 1D cross-field transport in a simple slab geometry. The coordinates are chosen so that the magnetic field is in the $\hat z$ direction and all gradients are in the $\hat x$ direction. Velocities are assumed to be in the perpendicular ($\hat x$ and $\hat y$) directions. This geometry is shown schematically in Figure~\ref{fig:geometrySchematic}. 
MITNS tracks and evolves the density, pressure, and velocity profiles of each particle species as well as evolving the magnetic field. 

\subsection{Density and Momentum}

For each species $s$, the evolution of the density $n_s$ is specified by a continuity equation 
\begin{gather}
\frac{\partial n_s}{\partial t} + \nabla \cdot \big( n_s \bv_s \big) = 0. \label{eqn:continuity}
\end{gather}
The ion velocities $\bv_i$ evolve according to the momentum equation:
\begin{align}
&\frac{\partial}{\partial t} (m_i n_i \bv_i) + \nabla p_i + \nabla \cdot (\pi_i + m_i n_i \bv_i \bv_i) \nonumber\\
&\hspace{60 pt}= Z_i e n_i ( \bE + \bv_i \times \bB ) + m_i n_i \sum_{s} \nu_{is} (\bv_s - \bv_i) + \mathbf{f}_{\text{th}, i} + m_i n_i \mathbf{g}. \label{eqn:ionMomentum}
\end{align}
Here $p_i$ is the scalar pressure, $\pi_i$ is the viscosity tensor, $Z_i$ is the ion charge state, $e$ is the elementary charge, $\bE$ is the electric field, $\bB$ is the magnetic field, $m_i$ is the ion mass, $\nu_{is}$ is the collision frequency of species $i$ with species $s$, $\mathbf{f}_{\text{th}, i}$ is the thermal friction force density, and $\mathbf{g}(t,x)$ is the gravitational acceleration. 

The collision frequency $\nu_{ab}$ for a species $a$ due to interactions with a species $b$ is given for any $a$ and $b$ (including ions and electrons) by \cite{Hinton}:
\begin{gather}
\nu_{ab} = \bigg( \frac{\sqrt{2} e^4 \log \Lambda}{12 \pi^{3/2} \epsilon_0^2}  \bigg) \bigg( Z_a^2 Z_b^2 \frac{m_a + m_b}{m_a^2 m_b} \bigg) \bigg( \frac{m_b T_a + m_a T_b}{m_a m_b} \bigg)^{-3/2} n_b \, , \label{eqn:collisionFrequency}
\end{gather}
where $\log \Lambda$ is the Coulomb logarithm and $T_s = p_s / n_s$ is the temperature of species $s$. 

The cross-field thermal force density on species $a$ can be written as \cite{Hirshman1981}:
\begin{align}
\mathbf{f}_{\text{th}, a} &= \sum_b \frac{3}{2} \frac{n_a \nu_{ab}}{\Omega_a} \frac{1}{1 + (m_a T_b / m_b T_a)} \bigg( \hat b \times \nabla T_a - \frac{Z_a}{Z_b} \frac{m_a}{m_b} \frac{T_b}{T_a} \hat b \times \nabla T_b \bigg), \label{eqn:NernstForceDensity}
\end{align}
where $\hat b$ is the unit vector in the direction of $\bB$. For systems with temperature gradients parallel to $\bB$, which are not considered here, there would be additional temperature-dependent force densities \cite{Braginskii1965}. 

The cross-field viscous forces in a magnetized plasma tend to be much smaller than the other forces, in addition to being quite complicated \cite{Braginskii1965}, especially in the low-flow case \cite{Simakov2003, Catto2004}.
However, the plasma cannot relax to the global thermodynamic equilibrium without the inclusion of some visosity to relax the flow shear.
Thus, we include only the multiple-species analog of the Braginskii $\eta_1$ component of the viscosity tensor,  which is both the simplest and most dominant viscous contribution in systems with geometric symmetries and very small $\hat x$-directed flows \cite{Kolmes2019}. 
In the case of a slab with all gradients in the $\hat x$ direction, the Braginskii viscous force density reduces to:
\begin{gather}
\nabla \cdot \pi_i = - \frac{\partial}{\partial x} \bigg( \frac{3 p_i \nu_{ii}}{10 \sqrt{2} \Omega_i^2} \frac{\partial v_{iy}}{\partial x} \bigg) \, \hat y , 
\end{gather}
where $\Omega_i \doteq Z_i e B / m_i$ is the ion gyrofrequency. 
The analogous expression for a multiple-ion plasma, as presented by Zhdanov \cite{Zhdanov}, can be written as 
\begin{gather}
\nabla \cdot \pi_a = - \frac{\partial}{\partial x} \bigg[ \frac{p_a}{4 \Omega_a^2} \sum_b \frac{\sqrt{2} m_a m_b \nu_{ab}}{(m_a + m_b)^2} \bigg( \frac{6}{5} \frac{m_b}{m_a} + 2 - \frac{4}{5} \frac{m_b}{m_a} \frac{Z_a}{Z_b} \bigg) \frac{\partial v_{ay}}{\partial x} \bigg] \, \hat y . 
\end{gather}
A more detailed discussion of viscosity, including the terms neglected by MITNS, can be found in \ref{sec:viscosity}. 

In the force balance equation for electrons, analogous to Eq.~(\ref{eqn:ionMomentum}), the small electron mass means that the inertial term and the electron viscosity can be ignored. In the slab geometry considered, dropping the inertial term is physically equivalent to the assumption that electron force balance is fast enough to be considered instantaneous. The resulting momentum equation is 
\begin{align}
0 = -e \bE - e \bv_e \times \bB - \frac{\nabla p_e}{n_e} + \sum_i m_e \nu_{ei} (\bv_i - \bv_e) + \frac{\mathbf{f}_{\text{th},e}}{n_e}, \label{eqn:electronMomentum}
\end{align}
which determines the electric field $\bE$. 

Thus, the continuity and momentum equations determine the evolution of the ion density and ion velocities, and self-consistently determine the electric field.

\subsection{Heat}

MITNS models the pressure evolution for each species $s$ by 
\begin{align}
&\frac{\partial}{\partial t} \bigg( \frac{3}{2} p_s \bigg) + \nabla \cdot \bigg(\mathbf{q}_{s \perp} + \frac{5}{2} p_s \mathbf{v}_s \bigg) = \mathbf{v}_s \cdot \nabla p_s + \sum_{s'} \frac{3 m_s n_s \nu_{ss'}}{m_s+m_{s'}} \big( T_{s'} - T_s \big) - \pi_s : \nabla \bv_s \nonumber \\
&\hspace{0 pt}+ \sum_{s'} \frac{m_s m_{s'} n_s \nu_{ss'}}{m_s + m_{s'}} \, (\bv_{s'} - \bv_s) \cdot \bigg( \bv_{s'} - \bv_s + \frac{3 \hat b}{2 Z_s Z_{s'} e B} \times \frac{Z_{s'} m_{s'} T_s \nabla T_s - Z_s m_s T_{s'} \nabla T_{s'}}{m_{s'} T_s + m_s T_{s'}} \bigg) .
\label{eqn:pressureEvolution}
\end{align}
The cross-field heat flux $\mathbf{q}_{s \perp}$ can be written as \cite{Hirshman1981}:
\begin{align}
\mathbf{q}_{s \perp} 
&= \frac{5 p_s}{2 m_s \Omega_s} \hat b \times \nabla T_s \nonumber \\
&\hspace{10 pt}+ \frac{p_s}{\Omega_s} \sum_{s'} \frac{\nu_{ss'}}{1 + (m_s T_{s'} / m_{s'} T_s)} \bigg\{ \frac{3}{2} (\bv_s - \bv_{s'}) \times \hat b\nonumber \\
&\hspace{40 pt}- \frac{m_{s'}}{m_s + m_{s'}} \bigg[ \bigg( \frac{13}{4} + 4 \, \frac{m_s T_{s'}}{m_{s'} T_s} + \frac{15}{2} \frac{m_s^2 T_{s'}^2}{m_{s'}^2 T_s^2} \bigg) \frac{\nabla_\perp T_s}{m_s \Omega_s} - \frac{27}{4} \frac{m_s}{m_{s'}} \frac{\nabla_\perp T_{s'}}{m_s \Omega_s} \bigg] \bigg\}. \label{eqn:heatFlux}
\end{align}
The physics behind this expression, including the appearance of the velocity terms, is discussed in greater detail in Section~\ref{sec:scaling}. 

A discussion of the viscous heating, and the approximations used by MITNS in modeling it, can be found in \ref{sec:viscosity}. Ultimately, MITNS models the viscous heating for ion species $i$ by:
\begin{align}
-\pi_i : \nabla \bv_i = \frac{p_i}{4 \Omega_i^2} \sum_s \frac{\sqrt{2} m_i m_s \nu_{is}}{(m_i + m_s)^2} \bigg( \frac{6}{5} \frac{m_s}{m_i} + 2 - \frac{4}{5} \frac{m_s}{m_i} \frac{Z_i}{Z_s} \bigg) \bigg( \frac{\partial v_{iy}}{\partial x} \bigg)^2. 
\end{align}
This expression works equally well for the case when $m_s$ and $m_i$ are comparable and the case when one is much larger than the other. 
The corresponding expression for electrons would be negligible due to the smallness of the electron-ion mass ratio, so it is not included in the code. 

The last term in Eq.~(\ref{eqn:pressureEvolution}) (written as a sum over $s'$) is the frictional heating. The total frictional heating due to interactions between species $s$ and $s'$ -- that is, including both the heating of $s$ due to collisions with $s'$ and the heating of $s'$ due to collisions with $s$ -- is determined by energy conservation \cite{Hazeltine}. The expression used here splits the frictional heating going into $s$ and that going into $s'$ so that each species receives a share that is inversely proportional to its mass. This is the simplest expression that satisfies energy conservation while also matching Braginskii's large-mass-ratio limit. Moreover, the dependence of the frictional heating on mass can be recovered by considering, e.g., the energy transfer associated with a binary collision between two particles of different masses. 

It is sometimes helpful to understand which of these terms are associated with reversible processes and which are associated with irreversible processes. In the absence of any external source terms (such as a particle source), the entropy production rate for species $s$ can be written as \cite{Hazeltine}
\begin{gather}
\Theta_s = \frac{W_s}{T_s} - \frac{\pi_s : \nabla \bv_s}{T_s} - \frac{\mathbf{q}_{s\perp}}{T_s} \cdot \frac{\nabla T_s}{T_s} \, ,
\end{gather}
where $W_s$ consists of the second and fourth terms of the RHS of Eq.~(\ref{eqn:pressureEvolution}) -- that is, the inter-species temperature equilibration and the frictional heating. Note that it is possible to have collisional particle transport without producing more than an infinitesimal amount of entropy. For instance, the collisional particle particle transport due to flow friction will be linear in $(v_{s'y}-v_{sy})$ whereas the associated heating and entropy production are quadratic in $(v_{s'y} - v_{sy})$, so sufficiently slow cross-field particle transport will be associated with vanishingly small time-integrated entropy production. 

\subsection{Maxwell Equations}

The remaining governing equations can be obtained from Maxwell's equations. The magnetic field evolves according to Faraday's law of induction: 
\begin{gather}
\frac{\partial \bB}{\partial t} = - \nabla \times \bE. \label{eqn:induction}
\end{gather}
Note that if all gradients are in the $\hat x$ direction, if $E_z = 0$, and if $\bB$ is initially in the $\hat z$ direction, then Eq.~(\ref{eqn:induction}) implies that $\bB$ will remain purely in the $\hat z$ direction for all time. 

The electron velocities can be determined from Amp\`ere's law. In a plasma where the Alfv\'en velocity $v_A$ is much smaller than the speed of light, the displacement current is an $\mathcal{O}(v_A^2/c^2)$ correction and can be neglected, so that Amp\'ere's law becomes: 
\begin{gather}
\nabla \times \bB = e \mu_0 \, \bigg( \sum_i Z_i n_i \bv_i - n_e \bv_e \bigg). \label{eqn:Ampere}
\end{gather}
Since the evolution of $\bB$ and $\bv_i$ are already determined, this can be used to obtain $\bv_e$. Moreover, since $\bB = B \hat z$ and all gradients are in the $\hat x$ direction, the $\hat x$ component of Eq.~(\ref{eqn:Ampere}) becomes 
\begin{gather}
v_{ex} = \frac{1}{n_e} \sum_i Z_i n_i v_{ix}. 
\end{gather}
This allows the electron continuity equation given by Eq.~(\ref{eqn:continuity}) to be rewritten as 
\begin{gather}
\frac{\partial}{\partial t} \bigg( n_e - \sum_i Z_i n_i \bigg) = 0. 
\end{gather}
If the plasma is initially quasineutral, then this can be replaced (for all times) with a simple quasineutrality condition: 
\begin{gather}
n_e = \sum_i Z_i n_i. \label{eqn:quasineutrality}
\end{gather}

Thus, Maxwell's equations and quasineutrality determine the evolution of the magnetic field, and self-consistently determine the electron density and velocity.

\section{Normalization and Dimensionless Parameters} \label{sec:dimensionless}

Physical parameters are normalized to characteristic values, such as the characteristic density $n_0$, temperature $T_0$, and magnetic field $B_0$. The ion mass is normalized to $m_p$, the proton mass. 
Define the characteristic proton thermal velocity and gyrofrequency by $v_{thp0} \doteq \sqrt{T_0 / m_p}$ and $\Omega_{p0} \doteq e B_0 / m_p$, respectively. 
Then define the following normalized quantities: 
\begin{gather}
\tn_s \doteq \frac{n_s}{n_0} \\
\tT_s \doteq \frac{T_s}{T_0} \\
\tilde{p}_s \doteq \frac{p_s}{n_0 T_0} \\
\tilde{t} \doteq \Omega_{p0} t \\
\frac{\partial}{\partial \tilde{x}} \doteq \frac{v_{thp0}}{\Omega_{p0}} \frac{\partial}{\partial x} \\
\tilde{\bv}_s \doteq \frac{\bv_s}{v_{thp0}} \\
\tilde{\bE} \doteq \frac{\bE}{v_{thp0} B_0} \\
\tB \doteq \frac{\bB}{B_0} \\
G(\tilde{t}, \tilde{x}) \doteq \frac{\mathbf{g}(t,x) \cdot \hat x}{\Omega_{p0} v_{thp0}} \, .
\end{gather}
$n_0$, $T_0$, $B_0$, $v_{thp0}$, and $\Omega_{p0}$ will not appear explicitly in the governing equations, except in a few combinations. These will be the physically relevant dimensionless parameters for the simulations. First, each species is associated with a gyrofrequency ratio 
\begin{gather}
W_s \doteq \frac{Z_s e B_0}{m_s \Omega_{p0}}
\end{gather}
(here and elsewhere, use $Z_s = -1$ for electrons). 

To evaluate the inverse Hall parameter $\nu_{ab} / \Omega_{p0}$ for species $a$ and $b$, it is convenient to decompose the inverse Hall parameter into a part which is a global constant for all species; a part which depends on the choice of species but not on any spatially local information; and a part that depends on local values of the densities and temperatures. As such, let 
\begin{gather}
C_0 \doteq \bigg( \frac{\sqrt{2} e^4 \log \Lambda}{12 \pi^{3/2} \epsilon_0^2}  \bigg) \bigg( \frac{n_0}{m_p^{1/2} T_0^{3/2} \Omega_{p0}} \bigg)
\end{gather}
and 
\begin{gather}
C_{ab} \doteq Z_a^2 Z_b^2 \sqrt{\frac{m_b}{m_a} \frac{m_p}{m_a+m_b}}
\end{gather}
so that, as per Eq.~(\ref{eqn:collisionFrequency}), 
\begin{gather}
\frac{\nu_{ab}}{\Omega_{p0}} = C_0 \, C_{ab} \, \bigg( \frac{m_b \tT_a + m_a \tT_b}{m_a + m_b} \bigg)^{-3/2} \tn_b .
\end{gather}

The last major dimensionless parameter, which appears in the nondimensional form of Amp\'ere's law, is defined by 
\begin{gather}
A \doteq \frac{B_0^2}{\mu_0 n_0 T_0}. 
\end{gather}
Physically, $A$ can be interpreted as twice the inverse plasma $\beta$, evaluated at the characteristic density, temperature, and magnetic field ($n_0$, $T_0$, and $B_0$, respectively), where $\beta$ is the ratio of the plasma pressure to the magnetic field energy density. 

The governing equations of the system can be rewritten in terms of these dimensionless quantities. The ion density evolution described by Eq.~(\ref{eqn:continuity}) becomes 
\begin{gather}
\frac{\partial \tn_i}{\partial \tilde{t}} = - \frac{\partial}{\partial \tilde{x}} \big( \tn_i \tilde{v}_{ix} \big). 
\end{gather}
The electron density is set instantaneously by Eq.~(\ref{eqn:quasineutrality}), which becomes 
\begin{gather}
\tn_e = \sum_i Z_i \tn_i. 
\end{gather}
Eq.~(\ref{eqn:ionMomentum}) defines the evolution of the ion momenta. Its $\hat x$ component can be written as 
\begin{align}
\frac{\partial \tv_{ix}}{\partial \tilde{t}} &= - \tv_{ix} \frac{\partial \tv_{ix}}{\partial \tilde{x}} + W_i (\tilde{E}_x + \tv_{iy} \tilde{B}) - \frac{m_p}{m_i} \frac{1}{\tn_i} \frac{\partial \tilde{p}_i}{\partial \tilde{x}} \nonumber \\
&\hspace{40 pt}+ \sum_s C_0 C_{is} \tn_s \bigg( \frac{m_s \tT_i + m_i \tT_s}{m_s + m_i} \bigg)^{-3/2} (\tv_{sx} - \tv_{ix}) + G
\end{align}
and its $\hat y$ component is 
\begin{align}
&\frac{\partial \tv_{iy}}{\partial \tilde{t}} = - \tv_{ix} \frac{\partial \tv_{iy}}{\partial \tilde{x}} + W_i (\tilde{E}_y + \tv_{ix} \tilde{B}) \nonumber \\
&+ \frac{\sqrt{2} C_0}{4 Z_i^2 \tn_i} \frac{m_i^2}{m_p^2} \frac{\partial}{\partial \tilde{x}} \bigg[ \frac{\tilde{p}_i}{\tB^2} \sum_s \frac{m_s m_p C_{is} \tn_s}{(m_i + m_s)^2} \bigg( \frac{m_s \tT_i + m_i \tT_s}{m_s + m_i} \bigg)^{-3/2} \bigg( \frac{6}{5} \frac{m_s}{m_i} + 2 - \frac{4}{5} \frac{m_s}{m_i} \frac{Z_i}{Z_s} \bigg) \frac{\partial \tv_{iy}}{\partial \tilde{x}} \bigg] \nonumber \\
&\hspace{0 pt}+ \sum_s C_0 C_{is} \tn_s \bigg( \frac{m_s \tT_i + m_i \tT_s}{m_s + m_i} \bigg)^{-3/2} \nonumber \\
&\hspace{50 pt} \times \bigg[ (\tv_{sy} - \tv_{iy}) + \frac{3}{2} \frac{1}{Z_i \tB} \frac{1}{1 + (m_i \tT_s / m_s \tT_i)} \bigg( \frac{\partial \tT_i}{\partial \tilde{x}} - \frac{Z_i m_i \tT_s}{Z_s m_s \tT_i} \frac{\partial \tT_s}{\partial \tilde{x}} \bigg) \bigg] . 
\end{align}
The electron velocities are set by Eq.~(\ref{eqn:Ampere}), which is 
\begin{gather}
\tv_{ex} = \frac{1}{\tn_e} \sum_i Z_i \tn_i \tv_{ix}
\end{gather}
and 
\begin{gather}
\tv_{ey} = \frac{A}{\tn_e} \frac{\partial \tB}{\partial \tilde{x}} + \frac{1}{\tn_e} \sum_i Z_i \tn_i \tv_{iy}. 
\end{gather}
Pressure evolution is set for all species by Eq.~(\ref{eqn:pressureEvolution}). This can be written as 
\begin{align}
\frac{\partial \tilde{p}_s}{\partial \tilde{t}} &= - \frac{5}{3} \frac{\partial}{\partial \tilde{x}} \big( \tilde{v}_{sx} \tilde{p}_s \big) + \frac{2}{3} \tilde{v}_{sx} \frac{\partial \tilde{p}_s}{\partial \tilde{x}} \nonumber \\
&\hspace{0 pt} + \frac{\partial}{\partial \tilde{x}} \frac{m_s / m_p}{Z_s \tB} \sum_{s'} \frac{\tilde{p}_s \tn_{s'}}{1+(m_s \tilde{T}_{s'}/m_{s'} \tilde{T}_s)} C_0 C_{ss'} \bigg( \frac{m_{s'} \tT_s + m_s \tT_{s'}}{m_s + m_{s'}} \bigg)^{-3/2} \nonumber \\
&\hspace{10 pt} \times \bigg\{ \tv_{sy} - \tv_{s'y} - \frac{m_{s'}}{m_s + m_{s'}} \frac{1}{Z_s \tB} \bigg[ \bigg( \frac{13}{6} + \frac{8}{3} \frac{m_s \tT_{s'}}{m_{s'} \tT_s} + 5 \frac{m_s^2 \tT_{s'}^2}{m_{s'}^2 \tT_s^2} \bigg) \frac{\partial \tT_s}{\partial \tilde{x}} - \frac{9}{2} \frac{m_s}{m_{s'}} \frac{\partial \tT_{s'}}{\partial \tilde{x}} \bigg] \bigg\} \nonumber \\
&\hspace{0 pt} + \sum_{s'} \frac{2 m_s \tn_s \tn_{s'}}{m_s + m_{s'}} C_0 C_{ss'} \bigg( \frac{m_{s'} \tT_s + m_s \tT_{s'}}{m_s + m_{s'}} \bigg)^{-3/2} \bigg\{ \tT_{s'} - \tT_s \nonumber \\
&\hspace{10 pt}+ \frac{1}{3} \frac{m_{s'}}{m_p} \bigg[ (\tv_{s'x} - \tv_{sx})^2 + (\tv_{s'y} - \tv_{sy})^2 + \frac{3 (\tv_{s'y} - \tv_{sy})}{2 Z_s Z_{s'} \tB} \frac{Z_{s'} m_{s'} \tT_s \partial_{\tilde{x}} \tT_s - Z_s m_s \tT_{s'} \partial_{\tilde{x}} \tT_{s'}}{m_{s'} \tT_s + m_s \tT_{s'}} \bigg] \bigg\}\nonumber \\
&\hspace{0 pt} + \frac{\sqrt{2}}{6} \frac{\tilde{p}_s}{Z_s^2 \tB^2} \frac{m_s^2}{m_p^2} \bigg( \frac{\partial \tv_{sy}}{\partial \tilde{x}} \bigg)^2 \nonumber \\
&\hspace{10 pt}\times \sum_{s'} \frac{m_s m_{s'} \tn_{s'}}{(m_s + m_{s'})^2} C_0 C_{ss'} \bigg( \frac{m_{s'} \tT_s + m_s \tT_{s'}}{m_s + m_{s'}} \bigg)^{-3/2} \bigg( \frac{6}{5} \frac{m_{s'}}{m_s} + 2 - \frac{4}{5} \frac{m_{s'}}{m_s} \frac{Z_s}{Z_{s'}} \bigg)  ,
\end{align}
where the final term (the viscous heating) is neglected for electrons. 
The magnetic field evolution, which is described by Eq.~(\ref{eqn:induction}), can be written as 
\begin{gather}
\frac{\partial \tB}{\partial \tilde{t}} = - \frac{\partial \tilde{E}_y}{\partial \tilde{x}}. 
\end{gather}
Finally, the electric field, which is set by  Eq.~(\ref{eqn:electronMomentum}), can be expressed as 
\begin{gather}
\tilde{E}_x = - \tv_{ey} \tB - \frac{1}{\tn_e} \frac{\partial \tilde{p}_e}{\partial \tilde{x}} + \frac{m_e}{m_p} \sum_i C_0 C_{ei} \tn_i \tT_e^{-3/2} (\tv_{ix} - \tv_{ex})
\end{gather}
and 
\begin{gather}
\tilde{E}_y = \tv_{ex} \tB + \frac{m_e}{m_p} \sum_i C_0 C_{ei} \tn_i \tT_e^{-3/2} \, \bigg[ (\tv_{iy} - \tv_{ey}) - \frac{3}{2} \frac{1}{\tB} \frac{\partial \tT_e}{\partial \tilde{x}} \bigg]. 
\end{gather}

\section{Tunable Physics and the Ettingshausen Effect} \label{sec:scaling}

It is often very useful to be able to turn on or turn off different physical effects in a simulation. MITNS includes a number of options to either turn off or to continuously scale down different physical effects (or to scale them  up). Most of these are quite simple. For instance, MITNS has a flag which can turn off temperature evolution, so that $T_s = T_0 = \text{constant}$ for all species. It also has a parameter that can scale the electron collisionality, which it accomplishes by sending $C_{ie} \rightarrow \alpha C_{ie}$ and $C_{ei} \rightarrow \alpha C_{ei}$ for all ion species $i$. This is particularly useful when studying physics that relies on the separation between the ion-ion and ion-electron collisional timescales. 
Similarly, it has a viscosity scaling parameter which sends $\nabla \cdot \pi_i \rightarrow \nabla \cdot \alpha \pi_i$ in the momentum equation and $-\pi_i : \nabla \bv_i \rightarrow -\alpha \pi_i : \nabla \bv_i$ in the heat equation. 

One feature of MITNS which is useful but physically nontrivial is the way in which it scales the thermal conductivity. When studying effects which deposit heat in different regions of the plasma, it is sometimes desirable to see what the temperature profiles would look like if the cross-field conductivity were reduced or removed. The collisional part of the particle flux $\Gamma_s$ and the heat flux $\mathbf{q}_s$ (not including the effects of viscosity) can be expressed as 
\begin{gather}
\begin{pmatrix}
\Gamma_s \\ \mathbf{q}_s
\end{pmatrix}_\text{collisional} = 
\begin{pmatrix}
\mathbf{A}_{11} & \mathbf{A}_{12} \\
\mathbf{A}_{21} & \mathbf{A}_{22}
\end{pmatrix}
\begin{pmatrix}
\{ \mathbf{v}_s - \mathbf{v}_{s'} \} \\
\{ \nabla T_{s'} \}
\end{pmatrix}, \label{eqn:transportMatrix}
\end{gather}
where the components $\mathbf{A}_{jk}$ are written as vectors because the collisional fluxes for species $s$ will depend on the velocities and temperature gradients of all species $s'$. When seeking to scale the cross-field conductivity by some factor $\alpha$, the most immediately intuitive solution would be to take 
\begin{gather}
\begin{pmatrix}
\mathbf{A}_{11} & \mathbf{A}_{12} \\
\mathbf{A}_{21} & \mathbf{A}_{22}
\end{pmatrix} \xrightarrow{?}
\begin{pmatrix}
\mathbf{A}_{11} & \mathbf{A}_{12} \\
\mathbf{A}_{21} & \alpha \mathbf{A}_{22}
\end{pmatrix}. 
\end{gather}
As it turns out, this is not the right approach, and in general the resulting system will be unstable. 

To see why, consider the physics of the Ettingshausen effect. 
The Ettingshausen effect is generally invoked to explain the appearance of $\bv_s - \bv_{s'}$ terms in the heat flux (that is, the $\mathbf{A}_{21}$ terms in the transport matrix) \cite{Braginskii1965, Hinton}. It provides the Onsager-symmetric heat flux to correspond with the thermal force. 
The effect follows from the dependence of collisionality on kinetic energy. Higher-energy particles tend to be less collisional, so if collisions are driving a particle flux, the flux will tend to preferentially move lower-energy particles. This results in a heat flux in the direction opposite that of the collisional particle flux. 

However, the collisional particle flux also has a component that depends on temperature gradients ($\mathbf{A}_{12}$). This results from thermal friction, and is also essentially the result of the temperature dependence of the collision frequency \cite{Braginskii1965, Hinton}. If the particle flux depends on $\nabla T_{s'}$ via thermal friction, and the heat flux has a part that depends on the particle flux via the Ettingshausen effect, then there is a temperature gradient-dependent heat flux (part of $\mathbf{A}_{22}$) that is not due to heat conduction at all but rather to energy-dependent particle fluxes that happen to be driven by temperature gradients. 

As the system approaches equilibrium, the net collisional particle fluxes will become small. The collisional cross-field flux for species $a$ can be obtained from the $\hat x$ component of Eq.~(\ref{eqn:ionMomentum}): 
\begin{gather}
\Gamma_s^\text{collisional} = n_s \sum_{s'} \frac{\nu_{ss'}}{\Omega_s} \bigg[ (v_{s'y} - v_{sy}) + \frac{3}{2 Z_s Z_{s'} e B} \frac{Z_{s'} m_{s'} T_s T_s' - Z_s m_s T_{s'} T_{s'}'}{m_{s'} T_s + m_s T_{s'}} \bigg] . \label{eqn:collisionalParticleFlux}
\end{gather}
It is sometimes convenient to denote the individual terms in this sum by $\Gamma_{ss'}$, so that $\Gamma_s^\text{collisional} = \sum_{s'} \Gamma_{ss'}$. 
If $\Gamma_s^\text{collisional} \approx 0$, then the velocity differences $(v_{s'y} - v_{sy})$ -- and, by extension, the collisional heat flux term $\mathbf{A}_{21} \cdot \{ \bv_s - \bv_{s'} \}$ in Eq.~(\ref{eqn:transportMatrix}) -- will be approximately proportional to some combination of the temperature gradients $\nabla T_s$. Depending on the particular scenario being simulated, this means that the $\mathbf{A}_{21} \cdot \{ \bv_s - \bv_{s'} \}$ heat flux can act either as a heat diffusion or as an anti-diffusion. If the coefficient $\mathbf{A}_{22}$ has been removed, then the system is missing its conventional cross-field thermal conductivity and the Ettingshausen $T_s'$ heat flux. The first of these is diffusive and the second tends to cancel the $\mathbf{A}_{21}$ heat flux when the system is close to equilibrium. If the $\mathbf{A}_{21}$ heat flux is anti-diffusive, and if $\mathbf{A}_{22}$ has been removed or sufficiently reduced, then the system will be unstable. 

In order to scale down the cross-field conductivity without making the system unstable, the solution is evidently to keep some $T_s'$-dependent heat flux from $\mathbf{A}_{22}$ in order to cancel the potentially antidiffusive contribution from $\mathbf{A}_{21}$. However, there is no single unambiguously correct way of splitting $\mathbf{A}_{22}$. One option would be to keep the part of $\mathbf{A}_{22}$ that comes from the Ettingshausen effect; one might expect a combined heat flux term that looks like $q_{s, \text{ Ettingshausen}} \sim T_s^2 \, \partial \Gamma_s / \partial T_s$ (since the Ettingshausen effect arises from the difference in fluxes for hotter and colder particles). The prospect of partitioning the heat conductivity based on the underlying physical mechanisms is appealing. However, this approach has a downside: the contributions to $\Gamma_s$ from the flow friction and the thermal friction scale differently with $T_s$, so $\partial \Gamma_{ss'} / \partial T_s$ does not necessarily vanish when $\Gamma_{ss'} \rightarrow 0$. 

An alternative approach -- and the one that is implemented in MITNS -- is instead to split $\mathbf{A}_{22}$ based on the criterion that the heat flux due to collisions between species $s$ and $s'$ should vanish when the corresponding particle flux $\Gamma_{ss'}$ does: 
\begin{align}
&q_{sx} \rightarrow - \frac{3 T_s}{2} \sum_{s'} \frac{\Gamma_{ss'}}{1+(m_s T_{s'} / m_{s'} T_s)} \nonumber \\
&\hspace{30 pt}- \alpha \cdot \frac{p_s}{\Omega_s} \sum_{s'} \frac{\nu_{ss'}}{1+(m_s T_{s'}/m_{s'} T_s)} \nonumber \\
&\hspace{60 pt} \times \bigg\{ \frac{m_{s'}}{m_s + m_{s'}} \bigg[ \bigg( \frac{13}{4} + 4 \, \frac{m_s T_{s'}}{m_{s'} T_s} + \frac{15}{2} \frac{m_s^2 T_{s'}^2}{m_{s'}^2 T_s^2} \bigg) \frac{\partial_x T_s}{m_s \Omega_s} - \frac{27}{4} \frac{m_s}{m_{s'}} \frac{\partial_x T_{s'}}{m_s \Omega_s} \bigg] \nonumber \\
&\hspace{90 pt} + \frac{9}{4} \frac{m_{s'} T_s}{m_{s'} T_s + m_s T_{s'}} \frac{\partial_x T_s}{m_s \Omega_s} - \frac{9}{4} \frac{Z_s}{Z_{s'}} \frac{m_s T_{s'}}{m_{s'} T_s + m_s T_{s'}} \frac{\partial_x T_{s'}}{m_s \Omega_s} \bigg\} . \label{eqn:splitFlux}
\end{align}
Eq.~(\ref{eqn:splitFlux}) reduces to the full heat flux given in Eq.~(\ref{eqn:heatFlux}) when $\alpha = 1$. When $\alpha \rightarrow 0$, each term in the sum vanishes when $\Gamma_{ss'}$ does. Moreover, the Onsager symmetry between $\mathbf{A}_{12}$ and $\mathbf{A}_{21}$ is preserved for any choice of $\alpha$. Eq.~(\ref{eqn:splitFlux}) is arguably the simplest possible expression with all of these properties.

\section{Numerical Implementation} \label{sec:implementation}

The current version of MITNS discretizes space using a uniform 1D grid. Some physical quantities ($n_s$, $v_{sy}$, $T_s$, $p_s$, $B$, and $E_x$) are tracked in the interior of each grid cell; others ($v_{sx}$ and $E_y$) are tracked on the edges. The electron density, electron velocity, electric field, and the temperatures can all be inferred at any given time from other quantities, so MITNS only needs to store and evolve the ion densities $n_i$, ion velocities $v_{ix}$ and $v_{iy}$, all species' pressures $p_s$, and the magnetic field $B$. As a result, for a simulation with $N_g$ grid cells and $N_i$ ion species, MITNS solves a coupled system of $(4 N_i + 2) N_g + N_i$ ODEs. 

When a cell-centered value is required for a quantity that is tracked on cell edges (or vice versa), the value is linearly interpolated from its two neighboring edges (or cells). Spatial derivatives are implemented using a centered second-order finite difference, where the derivative of a cell-centered quantity is defined as being edge-centered and vice versa. 
As a result, the system of ODEs has a banded structure, with the evolution of the dynamical variables in any given cell (or on any given edge) depending only on the values in their own cell (or on their own edge) and on the values in both the nearest-neighbor cells and the nearest-neighbor edges. 
Of course, the structure is slightly different for boundary cells and edges. The simulations enforce boundary conditions that do not allow flux through the top or bottom of the system, so $v_{sx}$ and the heat flux $q_{sx}$ vanish on the boundary edges. These boundary conditions have the advantage that they are physically simple. Moreover, they are reasonable approximations for a range of systems (for example, the outer liner in a compression experiment or, in a simplified limit, the boundary of a magnetically confined plasma). 

In the current version of the code, this no-flux condition is enforced by treating the system as mirror-symmetric at each boundary.
This means that the quantities $n_s$, $T_s$, and $B_z$ are symmetric at each boundary, while the quantities $v_{sx}$, $v_{sy}$, $E_y$, and $E_x$ are antisymmetric at each boundary.
These boundary symmetries must be handled differently for cell and edge centered values.
For cell-centered values, the first ghost cell must be equal to the boundary cell, while for edge-centered values, the first ghost edge must be equal to the second-to-last edge from the boundary.
The centering and boundary symmetries of each variable are listed in Table~\ref{tab:symmetries}.

\begin{table}[]
    \centering
    \begin{tabular}{|c|c|c|}
    \hline
        Variable & Centering & Boundary Symmetry \\
        \hline
        $n_s$ & Cell & Symmetric \\
        $T_s$ & Cell & Symmetric \\
        $v_{sx}$ & Edge & Antisymmetric \\
        $v_{sy}$ & Cell & Antisymmetric \\
        $E_x$ & Cell & Antisymmetric \\
        $E_y$ & Edge & Antisymmetic \\
        $B_z$ & Cell & Symmetric\\
        \hline
     \end{tabular}
    \caption{Centering and boundary symmetries for the various variables in MITNS. The electromagnetic variables can be understood by comparing to the field configuration for Yee's PIC scheme \cite{Yee1966}.}
    \label{tab:symmetries}
\end{table}

In order to evolve this system of coupled ODEs in time, MITNS can use any of three solvers. The first is a fourth-order Runge-Kutta solver; it is typically the slowest of the three, but its relative simplicity is sometimes convenient for benchmarking. The second is a variable-order, variable-timestep Adams-Moulton solver, using functional iteration for its nonlinear solve step. The third is a variable-order, variable-timestep Backwards Differentiation Formula solver, with Newton iteration for its nonlinear solve. 
The AM and BDF solvers both use implementations from the CVODE package \cite{Hindmarsh2005, Cohen1996}. 

\section{Sample Output and Performance Analysis} \label{sec:performance}
\begin{figure}
	\centering
	\includegraphics[trim={6.cm 0.4cm 8cm 1.0cm},clip, width=\linewidth]{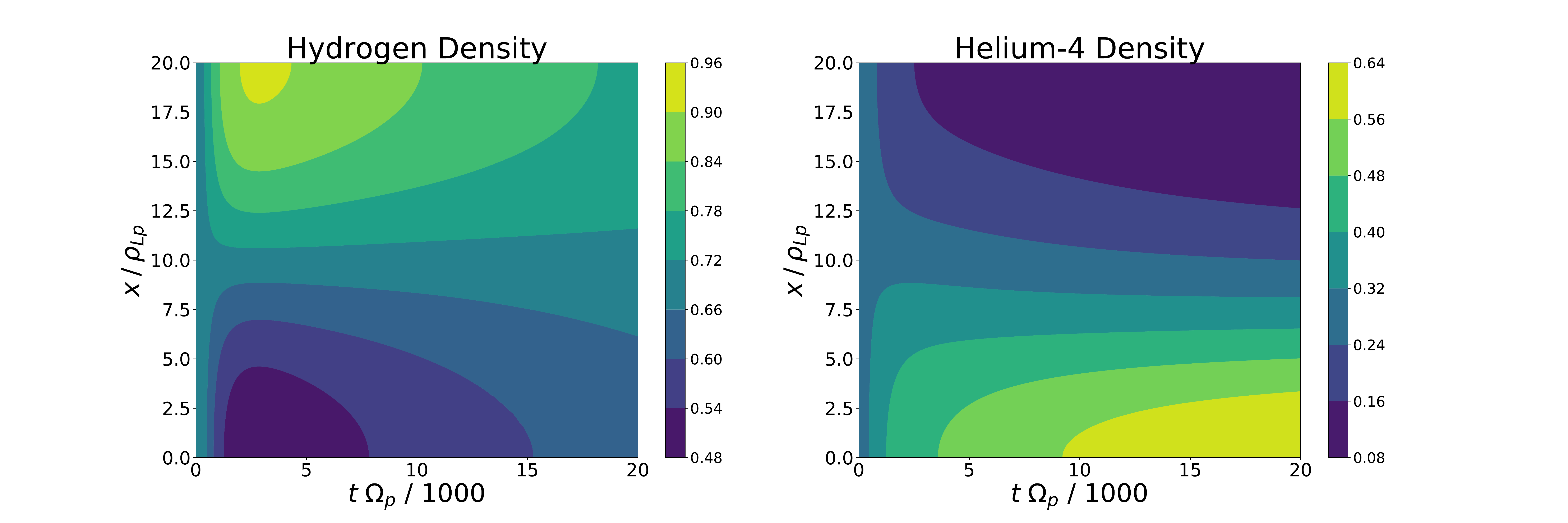}
	\caption{Density evolution of a plasma containing a mix of hydrogen and helium-4 in a gravitational potential. The time coordinate is normalized to one thousand proton gyroperiods and the spatial coordinate is normalized to a proton gyroradius (evaluated at the characteristic magnetic field $B_0$, density $n_0$, and temperature $T_0$ described earlier). This figure shows the relative motion of the ions described by Eq.~(\ref{eqn:impurityPinch}), where the ion species with the higher $m/Z$ initially falls in the potential while the species the the lower $m/Z$ initially rises. Later, the simulation begins to show both species fall as collisions between ions and electrons become important. This simulation used a ramp-up time of $300 \, \Omega_{p0}^{-1}$ for the potential. After that, the ions equilibrate with one another on a characteristic timescale that scales like $\nu_{ss'}^{-1} L^2 / \rho_{Ls}^2$. Electron-ion frictional equilibration takes about 80 times longer than ion-ion equilibration, so we don't see full electron-ion equilibrium here, but the system begins to move toward it. }
	\label{fig:pinchExample}
\end{figure}
\begin{figure}
	\centering
	\includegraphics[trim={0cm 0.5cm 0cm 0cm},clip,width=\linewidth]{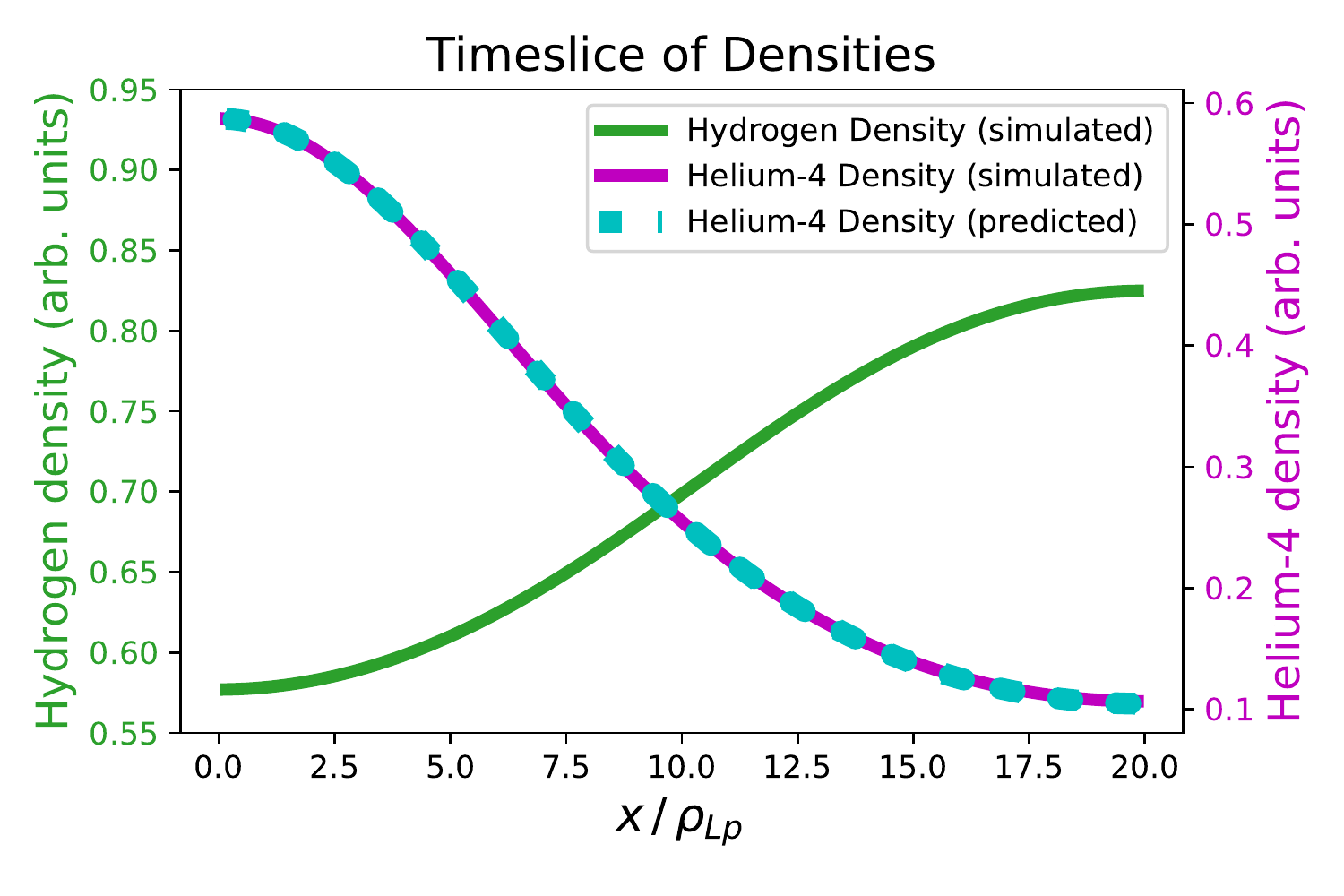}
	\caption{This figure shows how the ion density profiles at a particular timeslice ($t \approx 1200 \, \Omega_{p0}^{-1}$) correspond to the analytic prediction given by Eq.~(\ref{eqn:impurityPinch}). The green curve shows the simulated hydrogen profile. The magenta shows the simulated helium-4 profile. The dashed cyan curve shows the helium-4 profile that would be predicted by combining Eq.~(\ref{eqn:impurityPinch}) with the simulated hydrogen profile. The simulated and predicted helium-4 profiles show good agreement.} \label{fig:analyticCheck}
\end{figure}

One simple example that demonstrates some of the capabilities of MITNS, and which can be used to benchmark the performance of the code, is the accumulation of impurities in the presence of a mass-dependent potential.
In the limit where $\nabla T_s / T_s$ is small compared to $\nabla n_s / n_s$, different species’ density profiles are analytically expected \cite{Kolmes2018, Kolmes2020MaxEntropy} to satisfy
\begin{gather}
\bigg( n_a e^{\Phi_a/T_a} \bigg)^{1/Z_a} \propto \bigg( n_b e^{\Phi_b/T_b} \bigg)^{1/Z_b} , \label{eqn:impurityPinch}
\end{gather}
where $\Phi_s$ is the total potential applied to species $s$.
Consider a scenario in which an initially uniform plasma composed of hydrogen and helium-4 is subjected to a potential $\Phi_s(t,x)$ given by
\begin{gather}
\Phi_s(t,x) = - \frac{m_s g_0 L}{\pi} \tanh^4 \bigg( \frac{t}{t_\text{ramp}} \bigg) \cos \bigg( \frac{\pi x}{L} \bigg) \label{eqn:phiShape}
\end{gather}
for some potential strength parameter $g_0$ and ramp time $t_\text{ramp}$. The time dependence is chosen to be smooth and so that the potential will saturate after $t \approx t_\text{ramp}$. The behavior of the ion densities can be seen in Figure~\ref{fig:pinchExample}, with $t_\text{ramp} = 300 \, \Omega_{p0}^{-1}$ and $g_0 = \Omega_{p0} v_{thp0} / 100$.
Figure~\ref{fig:analyticCheck} shows the agreement between these profiles and the predictions from Eq.~(\ref{eqn:impurityPinch}).

\begin{figure}
    \centering
    \includegraphics[trim={1cm 0cm 2cm 0cm},clip,width=\linewidth]{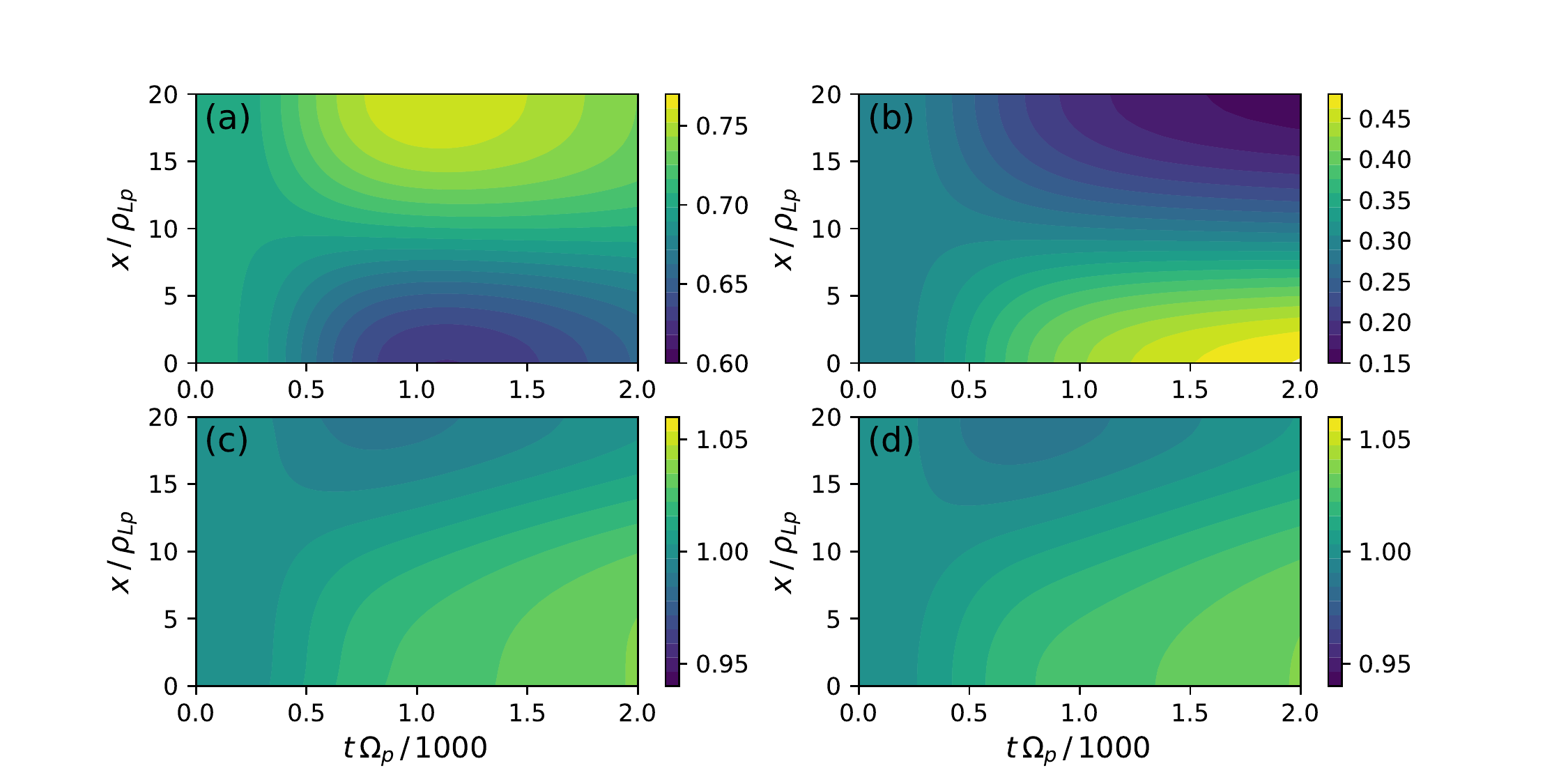}
    \caption{The evolution of a mixture of deuterium and tritium under a mass-dependent potential. The four panels are (a) the deuterium density, (b) the tritium density, (c) the deuterium temperature, and (d) the tritium temperature. This simulation includes the full classical heat conductivity. }
    \label{fig:kappaOn}
\end{figure}

\begin{figure}
    \centering
    \includegraphics[trim={1cm 0cm 2cm 0cm},clip,width=\linewidth]{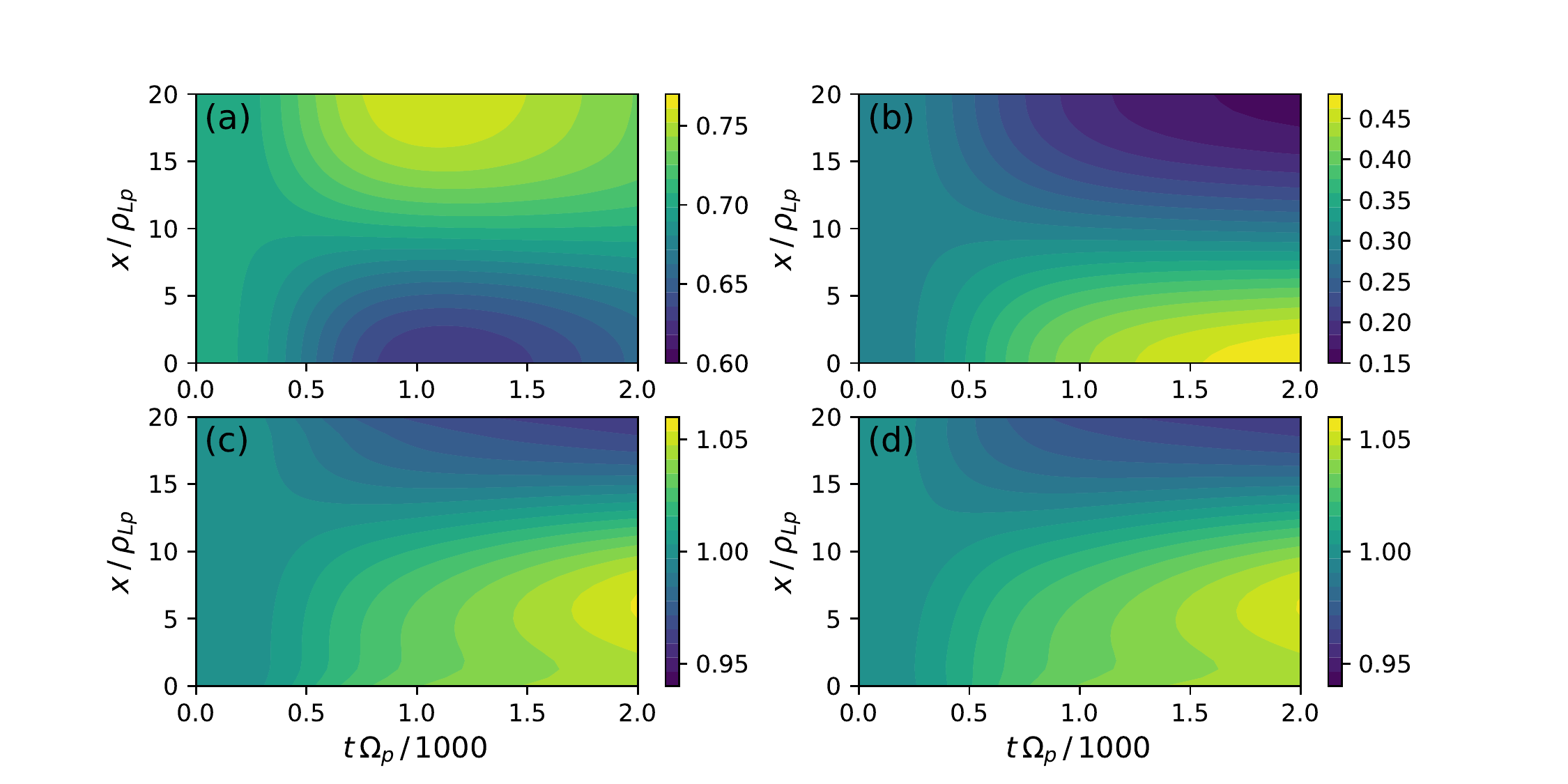}
    \caption{This simulation shows the same scenario as Figure~\ref{fig:kappaOn}, but with the classical heat conductivity suppressed. As before, the panels are (a) the deuterium density, (b) the tritium density, (c) the deuterium temperature, and (d) the tritium temperature. }
    \label{fig:kappaOff}
\end{figure}

Similar scenarios can showcase the tunable physics discussed in Section~\ref{sec:scaling}. For instance, consider an initially homogeneous mix of deuterium and tritium, with the same potential described in Eq.~(\ref{eqn:phiShape}) but with $g_0 = \Omega_{p0} v_{thp0} / 10$. Results with and without thermal conductivity are shown in Figures~\ref{fig:kappaOn} and \ref{fig:kappaOff}, respectively. The simulations without conductivity essentially show the spatial distribution of the heat source terms. Very similar MITNS simulations, both with and without heat conductivity, were used in Ref.~\cite{MlodikHeatPumparXiv}. Ref.~\cite{MlodikHeatPumparXiv} was a study of heat transport effects in rotating and compressing systems; applications included magnetized compression experiments like MagLIF. In that study, simulations without the heat conductivity were used to validate and illustrate analytic calculations that did not include the conductivity. Simulations that included heat conductivity made it possible to quantify the error associated with neglecting those terms. 


Simulations of scenarios like these -- with a mixture of two ion species and a potential given by Eq.~(\ref{eqn:phiShape}) -- can be used to benchmark the numerical performance of the code. The spatial and temporal discretization of the system of equations will each be associated with some numerical error. The error from the temporal evolution can be controlled with tolerance parameters passed to CVODE and is essentially independent from the implementation of MITNS itself. The error from the spatial discretization, on the other hand, is set by the second-order finite volume scheme described in Section~\ref{sec:implementation}.

To evaluate the code’s performance, we conducted simulations on increasingly fine grids in powers of two, from $N=4$ to $N = 512$.
Simulations were performed for a 70\% Hydrogen- 30\% Helium mix, in a system with $L/\rho_{Lp} = 20$, where $\rho_{Lp}$ is the characteristic proton Larmor radius. 
We then calculated both the estimated error and runtime associated with these simulations.

\begin{figure}
	\centering
	\includegraphics[width=0.8\linewidth]{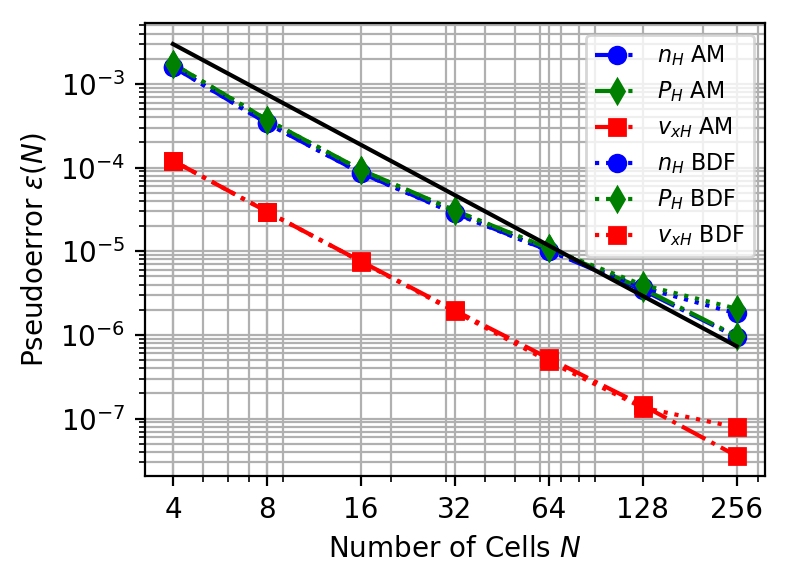}
	\caption{Pseudoerror vs. number of grid points in MITNS simulations, for several variables, for both Adams-Moulton and BDF integrators.
		The black line represents a scaling of $y\sim N^{-2}$.
		The pseudoerror initially scales as $\epsilon(N) \sim N^{-2}$, but this convergence slows around $N=64$.
		The slowing is more pronounced for the BDF method.} \label{fig:errorScaling}
\end{figure}

To estimate error, we calculated the pseudoerror, which does not require knowing the analytical solution to the problem.
To calculate the pseudoerror, the finest grid (in our case, $N=512$) is taken to represent the canonical solution; we then calculate error relative to these points.
To facilitate such analysis, the output of MITNS is edge-centered; thus, the spatial point $x_n$ associated with the $n$th gridpoint on a grid with $N$ cells (and $N+1$ edges) is the same as the spatial point $x_{2n}$ on a grid with $2N$ cells (and $2N+1$ edges).
Thus, every point on a coarse grid has a corresponding point on the finest grid.
For a function $y(t,x)$ on this grid, with numerical solutions $y_{t_i,x_j}$ and corresponding finest-grid solutions $Y_{t_i,x_i}$, the pseudoerror $\epsilon(N)$ is then given by:
\begin{align}
\epsilon(N) = \frac{1}{V} \sum_{i,j} \sqrt{\lp y_{t_i,x_j} - Y_{t_i,x_i} \rp^2}.
\end{align}
Here, $V$ is the total number of points in $t$ and $x$ that are summed over.
For the purpose of this analysis, we calculated the pseudoerror for the variables $n_H$, $P_H$, and $v_{xH}$.
The result of this pseudoerror analysis is shown in Fig.~\ref{fig:errorScaling}.
While both the Adams-Moulton and BDF schemes initially converge as $\epsilon(N) \sim N^{-2}$, as expected for a second-order scheme, they converge more slowly above $N = 64$.
This slowing convergence is particularly pronounced for the BDF method.

\begin{figure}
	\centering
	\includegraphics[width=0.7\linewidth]{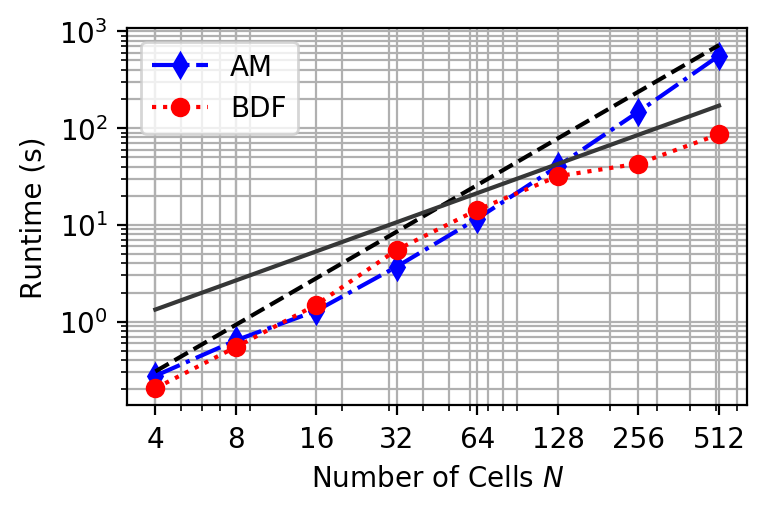}
	\caption{Runtime vs. grid size for each integration method.
		The black dashed line represents an $N^{1.6}$ scaling, while the dark gray solid line represents $N^1$ scaling.
		Although both methods initially scale as $N^{1.6}$, the BDF integrator runtime scaling becomes linear at $N=64$.} \label{fig:runtime}
\end{figure}

The corresponding runtime results (from a 2019 15'' Macbook Pro) for the same simulations are shown in Fig.~\ref{fig:runtime}.
The runtime initially increases as $T\sim N^{1.6}$ with the number of cells, with AM running slightly faster.
For large grids, however, the BDF runs much faster, scaling as $T\sim N$ after $N=64$.

\begin{figure}
	\centering
	\includegraphics[width=0.7\linewidth]{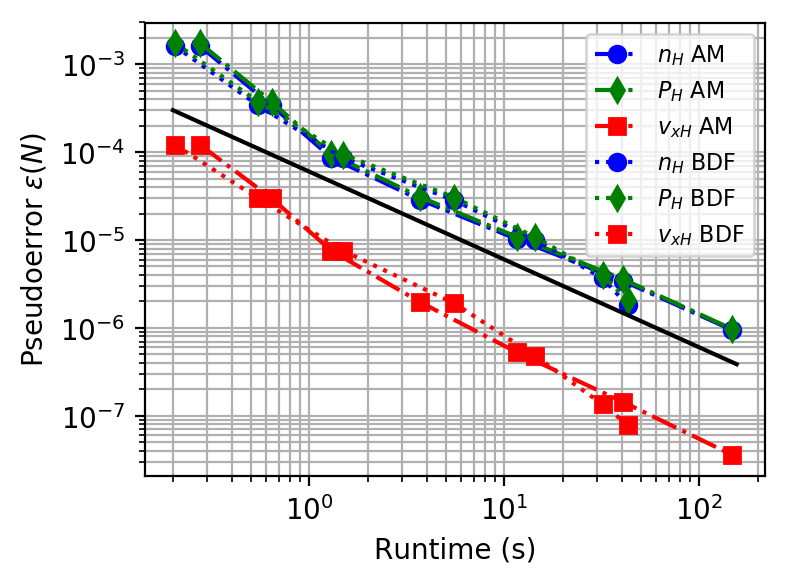}
	\caption{Combined data from Figs.~\ref{fig:errorScaling} and \ref{fig:runtime}, showing pseudoerror vs. runtime for each integration method. The black solid line represents a scaling of $T\sim N^{-1}$. Because the BDF method runs faster but with less accuracy than the AM method for large grids, the error vs. runtime is comparable for the two methods.} \label{fig:errorVruntime}
\end{figure}

For large grids, the BDF scheme runs faster but with higher error than the AM scheme.
It is thus natural to compare the error scaling with runtime for both methods, which is shown in Fig.~\ref{fig:errorVruntime}.
The relative speed of the BDF method for large grids is more pronounced than its relative increase in error, so that the BDF method has better error-vs-runtime performance.

\begin{figure}
	\centering
	\includegraphics[width=0.7\linewidth]{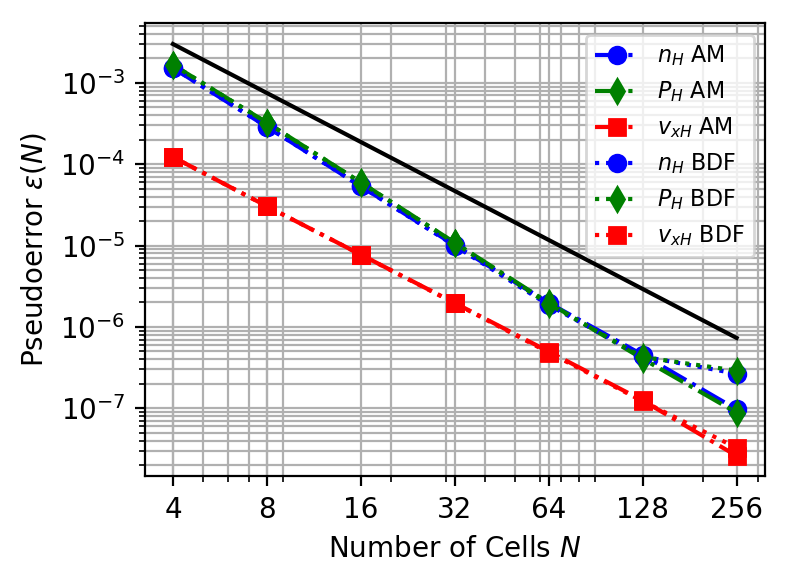}
	\caption{Pseudoerror vs. number of grid points in MITNS simulations, as in Fig.~\ref{fig:errorScaling}, but with viscosity and thermal conductivity turned off.
	The black line represents a scaling of $y\sim N^{-2}$.
	The pseudoerror is reduced by almost an order of magnitude relative to the case with the diffusive terms.} \label{fig:scalingNVNK}
\end{figure}

Interestingly, the deviation from $N^{-2}$ error convergence at large grid sizes appears to be related to diffusion-like terms, specifically the viscosity and thermal conductivity.
Since these are tunable parameters, we can turn them off; doing so results in $N^{-2}$ convergence to larger grid sizes (Fig.~\ref{fig:scalingNVNK}).
Exactly why the errors and runtimes scale this way at large $N$ is a matter of active research.
However, for the vast majority of problems, grid sizes $N \leq 64$ will be more than sufficient to get results to the desired accuracy.

\begin{figure}
	\centering
	\includegraphics[width=0.7\linewidth]{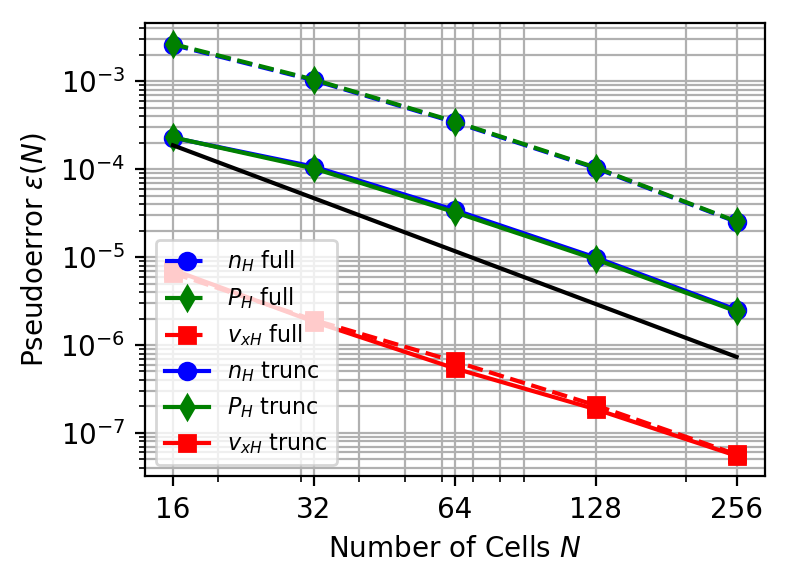}
	\caption{Pseudoerror vs. number of grid points in MITNS simulations, as in Fig.~\ref{fig:errorScaling}, but for a constant gravitational potential $\Phi_s \propto |x|$.
	Dashed lines show error considering all grid points, while solid lines exclude the two points closest to the boundaries.
	The error for $n$ and $P$ is much greater than the smooth case, due to the discontinuous forces at the boundaries, but this error is strongly concentrated in the boundary region.} \label{fig:scalingLinear}
\end{figure}

It is important to emphasize that the results we have shown are for a special class of potentials that are smooth and continuous at the boundaries, taking into account the periodicity of the potential.
When the potential is not smooth at each boundary, as for a constant gravitational field with potential $\Phi_s \propto |x|$, the analytic solution for the density and pressure will also not be smooth at the boundary, and so there will be error introduced by the reflection conditions for the ghost cells.
This leads to slightly slower runtimes and much greater error (Fig.~\ref{fig:scalingLinear}).
However, this error is strongly concentrated in the endpoints, so that the solution at other points remains robust. 

An alternate implementation of the boundary conditions (not currently included in the main version of the code), which linearly extrapolates the values of cell-centered variables at the outer-boundary edges from the neighboring inner values, seems to improve the error associated with non-periodic potentials. Of course, for practical purposes, it is already relatively straightforward to reduce the error to acceptable levels without having to go to an excessively large grid. 

\section{Discussion}

In its current form, MITNS is focused on a particular niche: the detailed fluid treatment of classical cross-field transport in a plasma with multiple ion species. The code is not designed to include the effects of turbulence or of transport in the direction parallel to $\bB$ (or, for that matter, transport in more than one of the perpendicular directions). It is also not designed to study the behavior of weakly magnetized or unmagetized systems. 
With that in mind, MITNS has already begun to be useful for problems within its purview \cite{MlodikHeatPumparXiv}. As far as we are aware, there is no established code in the field with the same capabilities. 
%

Future development is unlikely to change the focus of the code or its simple 1D geometry. However, there are a number of possible avenues for future improvement of the code's treatment of cross-field transport. 
One possibility would be to add the capability to simulate plasma undergoing compression. There are a number of laboratory experiments that involve compressing magnetized plasmas, and there is significant upside potential in understanding and controlling differential ion transport in these devices \cite{Ochs2018i, Ochs2018ii}. These upsides could include the control of fuel mix and impurities in fusion devices and the control of high-$Z$ species in compression devices used for X-ray generation. 

A similar extension would be to allow for a greater variety of boundary conditions and source terms. This could make it possible to model a greater variety of physical scenarios without losing the geometric simplicity of the current code. 

A third possibility would be to allow for transitions between charge states as well as neutral particles. This could be particularly important for plasma mass filters. These devices often rely on collisional transport to achieve species separation, and they tend to operate in regimes with significant populations of neutral and partially ionized particles \cite{Bonnevier1966, Lehnert1971, Krishnan1983, Zweben2018, Gueroult2018}. 

Finally, we could include the $\hat z$ component of the momentum equation.
This would involve adding many additional viscous and heating terms, but it would allow us to study transport in geometries with sheared flows parallel to the magnetic field.

\section*{Acknowledgements}

We would like to acknowledge Michael Mueller and Mikhail Mlodik for useful conversations. 
MITNS uses elements of CVODE/SUNDIALS, which are open source packages developed at Lawrence Livermore National Laboratory. 
This work was supported by NSF PHY-1805316, DOE DE-FG02-97ER25308, and NNSA 83228-10966 [Prime No. DOE (NNSA) DE-NA0003764]. 

\bibliographystyle{apsrev4-1} 
\bibliography{Master.bib}

\appendix
\section{Viscosity in a Simple Slab} \label{sec:viscosity}
Consider a slab geometry with all gradients in the $\hat x$ direction and all velocities in the $\hat x$ and $\hat y$ directions. Define 
\begin{gather}
W_{\alpha \beta} \doteq \frac{\partial v_\alpha}{\partial x_\beta} + \frac{\partial v_\beta}{\partial x_\alpha} - \frac{2}{3} \delta_{\alpha \beta} \nabla \cdot \bv, 
\end{gather}
where $\delta_{\alpha \beta}$ is the Kronecker delta. In this simple geometry, $W_{\alpha \beta}$ becomes 
\begin{gather}
W_{\alpha \beta} =
\frac{1}{3} \begin{pmatrix}
4 \, v_x' & 3 \, v_y' & 0 \\
3 \, v_y' & -2 \, v_x' & 0 \\
0 & 0 & -2 \, v_x'
\end{pmatrix}, 
\end{gather}
where $v_\alpha' = \partial v_\alpha / \partial x$. Then the Braginskii viscosity tensor \cite{Braginskii1965} is
\begin{gather}
\pi_{\alpha \beta} = \frac{1}{6} \begin{pmatrix}
- 2 \eta_0 \, v_x' - 6 \eta_1 \, v_x' - 6 \eta_3 \, v_y' & 
- 6 \eta_1 \, v_y' + 6 \eta_3 \, v_x' & 
0 \\
- 6 \eta_1 \, v_y' + 6 \eta_3 \, v_x' &
- 2 \eta_0 \, v_x' + 6 \eta_1 \, v_x' + 6 \eta_3 \, v_y' & 
0 \\
0 &
0 &
4 \eta_0 \, v_x'
\end{pmatrix}, 
\end{gather}
where for ions, to leading order in the inverse Hall parameter $\epsilon \doteq \nu_{ii} / \Omega_i$, 
\begin{gather}
\eta_0 = \frac{0.96 \sqrt{2} p_i}{\nu_{ii}} \\
\eta_1 = \frac{3}{10 \sqrt{2}} \frac{\nu_{ii} p_i}{\Omega_i^2} \\
\eta_3 = \frac{p_i}{2 \Omega_i} \, .
\end{gather}
To leading order, keeping in mind that $\eta_1 / \eta_3 \sim \eta_3 / \eta_0 \sim \epsilon$, the viscous force density in this system is 
\begin{gather}
\nabla \cdot \pi = - \hat x \, \frac{\partial}{\partial x} \bigg(  \frac{\eta_0}{3} \frac{\partial v_x}{\partial x} + \eta_3 \frac{\partial v_y}{\partial x} \bigg) - \hat y \, \frac{\partial}{\partial x} \bigg( \eta_1 \frac{\partial v_y}{\partial x} - \eta_3 \frac{\partial v_x}{\partial x} \bigg) . \label{eqn:viscousForceDensity}
\end{gather}
Braginskii's treatment was for a plasma with a single ion species. Zhdanov \cite{Zhdanov} gives the generalizations of these coefficients to a multiple-ion plasma. The expression for $\eta_{i3}$ is identical to the one found in Braginskii. $\eta_{i1}$ becomes 
\begin{gather}
\eta_{i1} = \frac{p_i}{4 \Omega_i^2} \sum_s \frac{\sqrt{2} m_i m_s \nu_{is}}{(m_i + m_s)^2} \bigg( \frac{6}{5} \frac{m_s}{m_i} + 2 - \frac{4}{5} \frac{m_s}{m_i} \frac{Z_i}{Z_s} \bigg). 
\end{gather}
The expression for the multiple-species $\eta_{i0}$ involves more complicated numerical coefficients, which are described in detail by Zhdanov \cite{Zhdanov}. However, like $\eta_{i1}$ and $\eta_{i3}$, the multiple-species form of $\eta_{i0}$ scales in essentially the same way (e.g. in the combined ion Hall parameter) as its single-species counterpart. 

Of the four terms in Eq.~(\ref{eqn:viscousForceDensity}), only the third is included in MITNS. It plays a qualitatively significant role over the longest timescales, since it prevents the system from fully relaxing until $v_y$ contains no shear. The other terms can reasonably be dropped. 

The first term in Eq.~(\ref{eqn:viscousForceDensity}) is negligible compared to the pressure force. Define $\tau_n$ as the characteristic timescale over which the ion density profiles evolve and define $\ell$ as the gradient scale length. The continuity equation implies that $v_{ix} / \ell \sim 1 / \tau_n$. Then 
\begin{gather}
\frac{\partial}{\partial x} \bigg( \frac{\eta_{i0}}{3} \frac{\partial v_x}{\partial x} \bigg) \sim \frac{p_i}{\ell \tau_n \nu_{ii}}
\end{gather}
whereas 
\begin{gather}
\frac{\partial p_i}{\partial x} \sim \frac{p_i}{\ell} \, .
\end{gather}
In other words, this part of the viscosity is negligible so long as the ions collide many times over the timescale $\tau_n$. This assumption is already necessary in order to use a high-collisionality closure. 

The second term in Eq.~(\ref{eqn:viscousForceDensity}) can be ignored for similar reasons. If $v_{thi}$ is the characteristic thermal velocity of species $i$ and $\rho_{Li}$ is the characteristic Larmor radius, 
\begin{gather}
\frac{\partial}{\partial x} \bigg( \eta_{i3} \frac{\partial v_{iy}}{\partial x} \bigg) \sim \frac{p_i}{\ell} \frac{v_{iy}}{v_{thi}} \frac{\rho_{Li}}{\ell} \, .
\end{gather}
Barring extraordinarily fast flows in the $\hat y$ direction, this will be small compared to the pressure force density. 

The fourth term in Eq.~(\ref{eqn:viscousForceDensity}) is also small compared to the pressure force density (even smaller than the first term was), but since it is oriented in the $\hat y$ direction, it is most useful to compare it with another term in the $\hat y$ momentum equation. This part of the viscous force density scales like
\begin{gather}
\frac{\partial}{\partial x} \bigg( \eta_{i3} \frac{\partial v_{ix}}{\partial x} \bigg) \sim \frac{p_i v_{ix}}{\Omega_i \ell^2}
\end{gather}
while the corresponding component of the $\bv \times \bB$ force scales like 
\begin{gather}
Z_i e n_i v_{ix} B \sim \frac{p_i v_{ix}}{\Omega_i \rho_{Li}^2} \, .
\end{gather}
Again, the viscous term in question will be comparatively small. Moreover, neither this nor either of the other two terms in Eq.~(\ref{eqn:viscousForceDensity}) not included in MITNS have the same kind of qualitative importance that the third term does. 

The leading-order viscous heating for species $i$ is 
\begin{align}
Q_\text{visc} &= - \pi_i : \nabla \bv_i \\
&= \frac{\eta_{i0}}{3} \bigg( \frac{\partial v_{ix}}{\partial x} \bigg)^2 + \eta_{i1} \bigg( \frac{\partial v_{iy}}{\partial x} \bigg)^2 .\label{eqn:Qvisc}
\end{align}
Using the continuity equation and defining $\tau_n$ in the same way as before, 
\begin{gather}
\frac{\eta_{i0}}{3} \bigg( \frac{\partial v_x}{\partial x} \bigg)^2 \sim \frac{p_i}{\nu_{ii} \tau_n^2} \, .
\end{gather}
Meanwhile, the compressional heating scales like 
\begin{gather}
Q_\text{compressional} \sim \frac{p_i}{\tau_n} \, .
\end{gather}
So long as $\nu_{ii} \tau_{n} \gg 1$, the $v_x$-dependent term in Eq.~(\ref{eqn:Qvisc}) can be neglected. The $v_y$-dependent term will often also be small, but it is less clear that it will be small in all cases, so it is included in the code. 

Of course, there is no reason why the code could not also include sub-dominant terms that we do not expect to be important. Indeed, future versions may do so. But there are some advantages in a simpler system of equations: they make the code easier to implement and easier to test, and they make the code's physics output more straightforward to understand. 

\end{document}